\documentclass[12pt]{iopart}
\usepackage{color}
\usepackage{graphicx}% Include figure files
\usepackage{iopams}% Align table columns on decimal point

\begin{document}

\hyphenation{acce-le-ra-tion tomo-graphy prio-ri maxi-mum}
\newcommand{\vect}[1]{\bi{#1}}
\newcommand{\gerade}[0]{\textit{gerade} }
\newcommand{\ungerade}[0]{\textit{ungerade} }

\title[Short laser pulses for molecular orbital tomography]{Control of recollision wave packets for molecular orbital tomography using short laser pulses}% Force line breaks with \\

\author{Elmar V van der Zwan and Manfred Lein}
 
\address{
Institute for Physics and Center for Interdisciplinary Nanostructure Science and Technology, University of Kassel, Heinrich-Plett-Stra{\ss}e 40, 34132 Kassel, Germany
}%
\ead{zwan@physik.uni-kassel.de}

\begin{abstract}
The tomographic imaging of arbitrary molecular orbitals via high-order harmonic generation requires that electrons recollide from one direction only. Within a semi-classical model, we show that extremely short phase-stabilized laser pulses offer control over the momentum distribution of the returning electrons. By adjusting the carrier-envelope phase, recollisions can be forced to occur from mainly one side, while retaining a broad energy spectrum. The signatures of the semi-classical distributions are observed in harmonic spectra obtained by numerical solution of the time-dependent Schr\"{o}dinger equation.
\end{abstract}


\maketitle
A system driven by a strong laser field emits high-frequency radiation in a process called high-order harmonic generation (HHG). For atoms and molecules, HHG is well understood in terms of the three-step picture \cite{Corkum93} as a sequence of field ionization, acceleration of the free electron in the laser field and recollision with the core. In recent years it has both become possible to use HHG to generate attosecond XUV pulses \cite{Paul01,Mairesse03,Tzallas03,Hentschel01,Drescher01}, and to study molecular properties \cite{Lein02,Kanai05,Baker06,Lein07}. Since the ground-breaking work of Itatani \etal \cite{Itatani04} it is known that HHG can be used to directly image molecular electronic orbitals in a scheme known as molecular orbital tomography. Itatani \etal approximately reconstructed the highest occupied electronic orbital of N$_2$ including its phase by comparing the high-harmonic spectra for different orientations of the molecule in the laser field with the spectrum of a reference system, namely atomic Ar, and combining this with knowledge about the electronic orbital of Ar. 

There are two main assumptions that enable us to perform the reconstruction of the molecular orbital; the first is that the returning electronic wave packets for the atomic and molecular case are very similar. This assumption stems from the fact that in the three-step picture the first step, tunnelling ionization, depends mainly on the instantaneous field strength and ionization energy. The two ionization energies are very similar. Also the second step, free electron acceleration, is identical for the two systems molecular N$_2$ and atomic Ar if the binding potential is neglected. The second assumption is that the returning electrons in the third step of the three-step picture can be described by plane waves. This is supported by the fact that the classical trajectories leading to the higher harmonics involve spending roughly half a laser period in the continuum. This is enough time for the wave packet to spread out considerably to a diameter much larger than that of the core, leading to the effective plane-wave character. In reality, the incoming wave packet is distorted by the core potential, which will be a source of errors for the procedure. Due to the plane-wave approximation, the matrix element describing the recombination process leading to high-harmonic generation becomes proportional to a Fourier transform of the molecular orbital. The spectra for many different orientations of the molecule in the laser field and information about the continuum wave packet at the moment of recombination, obtained from the atomic reference system, can then be combined to reconstruct a projection of the molecular orbital onto the plane perpendicular to the the laser propagation axis. 

In general, recolliding electrons with both positive and negative momenta contribute to the generation of the same harmonic generation, each through their own recombination matrix element. For an arbitrary orbital, referring here to an orbital that does not have \gerade or \ungerade symmetry, there is no obvious way to deduct the matrix element needed for the reconstruction from knowledge about the harmonic spectrum and the continuum wave packet. The reconstruction of an arbitrary orbital therefore requires that the wave packets approach the core from one side only \cite{vanderZwan07, vanderZwan07-2}. We propose achieving this behaviour of the continuum wave packet by using extremely short tailored laser pulses for high-harmonic generation. By stabilization and control of the carrier-envelope phase (CEP), i.e.\ the phase between the carrier wave and the envelope of an extremely short pulse, the returning wave packet can be tailored. Because of the very-few-cycle character of the pulse, the CEP can be chosen such that classical trajectories in the unwanted direction are effectively suppressed. 

\section*{Return probability}

To characterize the returning wave packet we calculate semi-classically the probability that an electron returns with momentum $k$. We sample the ionization time with 3000 points per laser cycle. At each ionization time $t_{\mathrm{i}}$, the electron tunnels with tunnelling probability (atomic units are used throughout)
\begin{equation}
P_{\mathrm{I}}\sim \exp\left(-\frac{4 I_{\mathrm{p}} \sqrt{2 I_{\mathrm{p}}}}{3 |E(t_{\mathrm{i}})|} \right),     
\end{equation}
where $E(t_{\mathrm{i}}$) is the laser electric field at the moment of ionization and $I_{\mathrm{p}}$ is the ionization energy. After tunnelling, the electron follows a classical trajectory, starting with zero velocity at position zero. Every classical return receives a weight based on the tunnelling probability and a factor $\tau^{-3}$, where $\tau$ is the time the electron spends in the continuum until the time of return. This factor reflects the effect of wave-packet spreading and comes from the Lewenstein model \cite{Lewenstein94}. We then add the contributions from the different trajectories to arrive at the probability that the electron returns to the core with a momentum $k$ in the range $k_{\mathrm{i}}-\Delta k / 2 < k \le k_{\mathrm{i}}+\Delta k / 2$, where $k_{\mathrm{i}}$ is the momentum of the i'th momentum bin and $\Delta k$ is the width of a bin. In contrast with the full Lewenstein model, interferences between the different paths leading to the same harmonic are not included in this model. This interference is irrelevant for the purpose of achieving one-sided electron returns.

We consider two types of analytical representations for extremely short and linearly-polarized laser pulses. The first type are 3-cycle pulses with a $\sin^2$-envelope
\begin{equation}
E_s(t) = -E_0 \sin^2 \left(\frac{\pi t}{3 T}\right) \cos(\omega t + \phi),
\end{equation}
where $E_0$ is the maximum field amplitude, $T$ is the laser period and $\phi$ is the CEP. We choose a laser wavelength of 780 nm and an intensity of $5\cdot 10^{14}$ W/cm$^2$ for both pulses. The length of such a pulse in terms of the full width at half maximum of the intensity (FWHM) is 2.84 fs. The second type are pulses with a Gaussian envelope. These pulses are in principle infinitely long but for the semi-classical calculation we consider only the centre 4 cycles of the pulses
\begin{equation}
E_g(t) = E_0 \exp \left(-1.5 \frac{(t-2 T)^2}{T^2} \right) \cos(\omega t + \phi).
\end{equation}
These pulses have a FWHM of 2.50 fs. Both mentioned values of the FWHM are comparable to one laser cycle (2.6 fs). The choice of simple analytical expressions for the time-dependent field strength leads to small but unimportant violations of the rule $\int E(t) \rmd t=0$ for the pulses with a Gaussian envelope. 

\section*{Results}

\begin{figure}[htbp]
  \centering
  \includegraphics[height=0.45\paperheight, angle=-90]{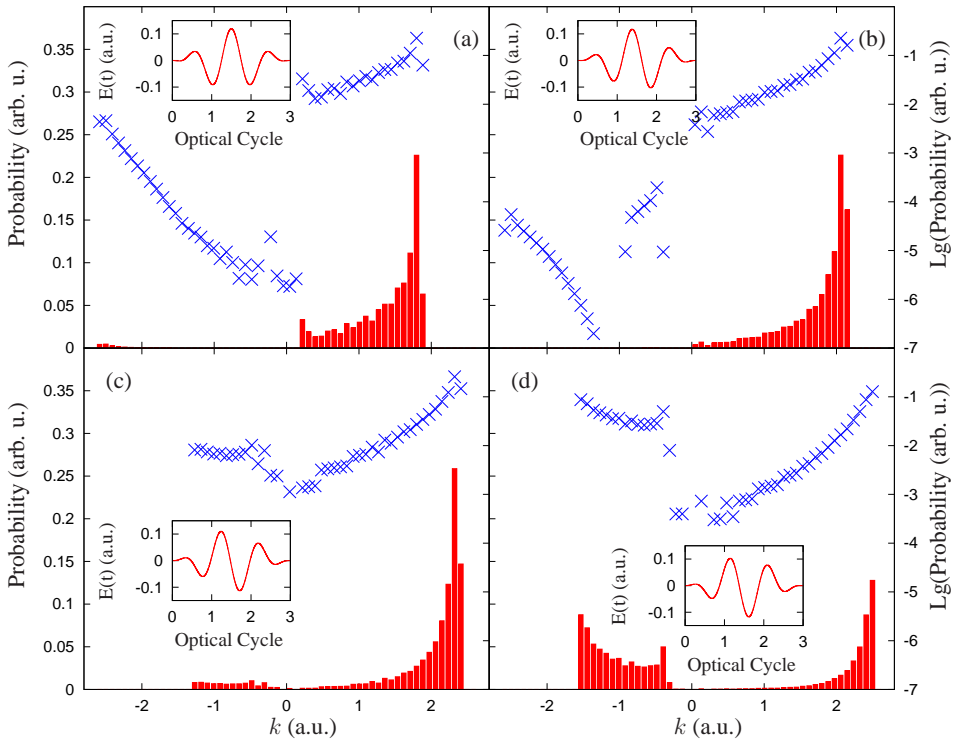}
  \caption{(colour online). Returning electron momentum distributions for the 3-cycle $\sin^2$-pulses. \mbox{(a) CEP $\phi=0$ rad}, \mbox{(b) $\phi=0.25\pi$ rad}, \mbox{(c) $\phi=0.55\pi$ rad} and \mbox{(d) $\phi=0.75\pi$ rad}. The distributions are shown on a linear scale (bars), and on a logarithmic scale (crosses). }
  \label{fig: 3cyclecos2}
\end{figure}
\begin{figure}[htbp]
  \centering
  \includegraphics[height=0.45\paperheight, angle=-90]{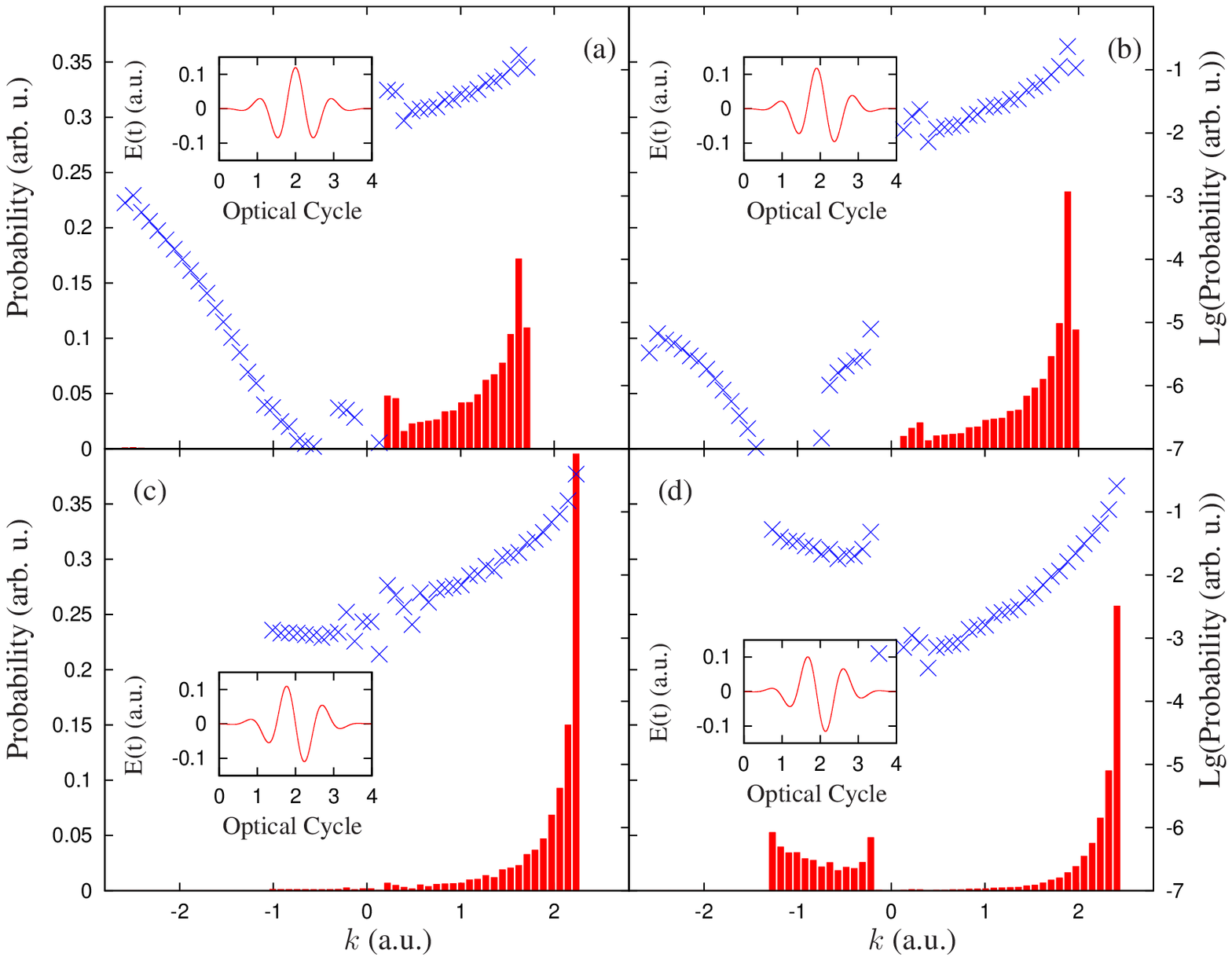}
  \caption{(colour online). Returning electron momentum distributions for the Gaussian pulses. \mbox{(a) CEP $\phi=0$ rad}, \mbox{(b) $\phi=0.2\pi$ rad}, \mbox{(c) $\phi=0.5\pi$ rad} and \mbox{(d) $\phi=0.7\pi$ rad}. The distributions are shown on a linear scale (bars), and on a logarithmic scale (crosses).}
  \label{fig: 4cyclegaussian}
\end{figure}
\begin{figure}[htb]
  \centering
  \includegraphics[width=0.4\paperwidth, angle=0]{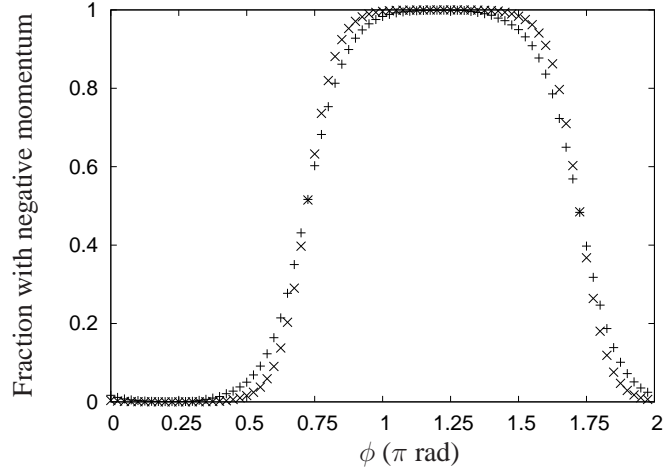}
  \caption{Fraction of electrons returning with negative momentum as a function of the CEP. Plus symbols correspond to the 3-cycle $\sin^2$-pulse, crosses to the Gaussian pulse.}
  \label{fig: leftsidefraction}
\end{figure}

We set the ionization potential equal to $I_{\mathrm{p}} = 30.2$ eV, in accordance with the ionization potential of a frequently used 2D model system for H$_2^+$ \cite{Lein02}. In figures \ref{fig: 3cyclecos2} and \ref{fig: 4cyclegaussian} the probability that an electron returns with momentum $k$ is depicted for the two different pulse types and different CEPs. The bars reflect that probability on a linear scale, the crosses on a logarithmic scale. For each pulse the sum of the probabilities is normalized to 1. The insets show the laser pulse used to generate the data. Subjected to an infinitely long pulse of the mentioned intensity and wavelength, the fastest returning electrons would have an energy of 3.17 $U_{\mathrm{p}}$ corresponding to $k\simeq \pm 2.57$ au, where $U_{\mathrm{p}}=E_0^2/(4\omega^2)$ is the ponderomotive potential \cite{Corkum93}. The figures show that the CEP of an extremely short pulse has a very large influence on the characteristics of the returning wave packet, and that tuning of the CEP allows one to select the characteristics that one desires. This is also apparent in figure \ref{fig: leftsidefraction}, which shows the fraction of the electrons returning with negative momentum as a function of the CEP for the two pulses. For $\phi=0$ rad, the pulses have their main ionizing half-cycle in the positive direction, which means that the electrons leave towards the negative direction. As a consequence, most electron returns occur with positive momentum then. 

An optimal pulse for use with the molecular orbital tomography should meet two requirements: (i) all returning electrons should come from the same side, i.e. zero probability for electrons returning with the opposite sign of the momentum; \mbox{(ii) the} energy spectrum of the returning electrons should be broad, so that the Fourier transform of the orbital can be accurately extracted from the HHG spectra \cite{vanderZwan07}. The results show that a relatively large range of CEPs that are suitable for molecular orbital tomography can be found for both pulses. For both pulses the results are very similar. This indicates that many types of phase-stabilized extremely short laser pulses will be suitable for the procedure. These pulses should have an ionizing half-cycle where the amplitude of the electric field is significantly higher than everything earlier, and the following half-cycle in the opposite direction should be comparable to the ionizing half-cycle in strength, to force the electrons to return. Everything after the `returning'-half-cycle should be significantly lower in amplitude, otherwise electrons will return from the other side with a significant probability. The reason that these effects can be achieved with the few-cycle pulses proposed in this paper, is that the ionization rate depends exponentially on the field amplitude. A nice demonstration that small field-amplitude changes can have large influence on harmonic spectra can be found in \cite{Cao07}. Because of this effect, we expect no problems with experimental pulses that have more smeared out tails, or with pre- and postpulses. For a different $I_\mathrm{p}$ the qualitative behavior will be the same, as $I_\mathrm{p}$ has no influence on the classical trajectories. Small differences in relative ionization rates might lead to a slightly different CEP being optimal for a different $I_\mathrm{p}$ though.  

\section*{Harmonic spectra}

\begin{figure}[htb]
  \centering
  \includegraphics[width=0.4\paperwidth, angle=0]{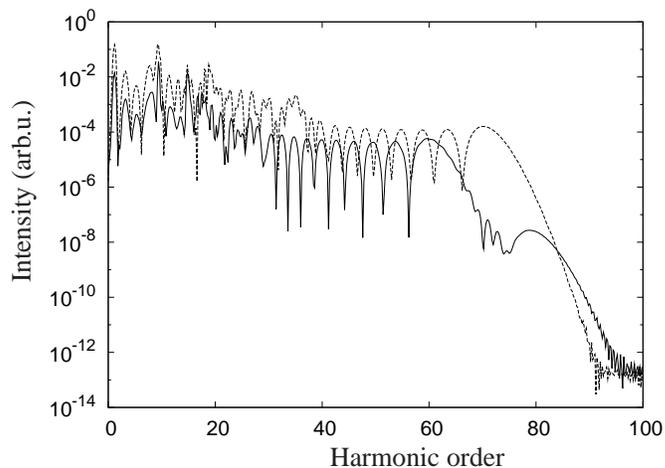}
  \caption{Harmonic spectra calculated using 3-cycle $\sin^2$-pulses with a CEP of $\phi=0.25\pi$ rad (solid line) and $\phi=0.55\pi$ rad (dashed line). The latter spectrum was multiplied with a factor 10 to separate the two curves graphically.}
  \label{fig: spectra}
\end{figure}

A quantum-mechanical signature of the above-mentioned effects can be found in the high-harmonic spectra generated with these pulses. For this purpose we solve the time-dependent Schr\"{o}dinger equation numerically using the split-operator method for a 2D He$^+$ model ion. We use the soft-core potential 
\begin{equation}
V(\vect{r})=-\frac{2}{\sqrt{\vect{r}^2+a^2}}
\end{equation}
with the softness parameter $a^2$ adjusted such that \mbox{$I_{\mathrm{p}}=30.2$ eV}. In figure \ref{fig: spectra} harmonic spectra associated with two values of the CEP for the 3-cycle $\sin^2$-pulse are shown. To ensure convergence, the wave function has been propagated for two more cycles after the end of the pulse. In both spectra, the low-frequency part is rather irregular due to the interference of several trajectories, while the high-frequency part shows the typical regular pattern of interference between short and long trajectories \cite{Milosevic02}. The latter is the pattern that we expect for uni-directional recollisions. 

For the case of $\phi=0.25\pi$ rad, the transition to the regular interference pattern occurs at roughly the 28th harmonic. To relate the energy of an emitted photon $\Omega$ with the momentum of the returning electron $k$, we use the energy-conserving dispersion relationship $\Omega=k^2/2+I_{\mathrm{p}}$. The 28th harmonic then corresponds to $k\simeq \pm 1$ au. As can be seen in figure \ref{fig: 3cyclecos2}, $|k|\simeq 1$ au is exactly the limit above which contributions come only from electrons with positive momentum. Also the cutoff at the 60th harmonic can be related to the classical picture, corresponding to the cutoff of $k\simeq 2.2$ au for electrons with positive momentum. The second cutoff comes from the less-likely events involving electrons with negative momentum, and extends clearly a couple of harmonics beyond the classically calculated and expected cutoff at the 76th harmonic, corresponding to $k\simeq -2.57$ au. 

In the case of $\phi=0.55\pi$ rad, the classical picture suggests a single cutoff at $k\simeq 2.4$ au, corresponding to the 68th harmonic. This is exactly what we see in figure \ref{fig: 3cyclecos2}. However, there is a small discrepancy between the TDSE and classical result concerning the location of the transition to a more regular spectrum. In the classical calculation, the fastest electrons with negative momentum have $k\simeq -1.3$ au. This would correspond to a transition in the spectrum around the 33th harmonic. The TDSE result shows this transition around the 37th harmonic, which may be due to modifications of the electron trajectories by the Coulombic potential. 

\section*{Conclusion}

A necessary condition for accurate tomographic reconstruction of arbitrary molecular orbitals is that all recombining electrons return from the same direction. We have shown that this is achieved by using extremely short phase-stabilized laser pulses. Because of the exponential dependence of the tunnelling rate on the field amplitude, the choice of envelope shape is not very critical.

\ack

The authors wish to thank C.C. Chiril\u{a} for fruitful discussions and wish to acknowledge discussions within the NSERC SRO network ``Controlled electron re-scattering: femtosecond, sub-angstrom imaging of single molecules''. 	

\section*{References} 

\bibliographystyle{unsrt}
\bibliography{references}

\begin{thebibliography}{10}

\bibitem{Corkum93}
P.~B. Corkum.
\newblock Plasma perspective on strong-field multiphoton ionization.
\newblock {\em Phys. Rev. Lett.}, 71(13):1994, 1993.

\bibitem{Paul01}
P.~M. Paul, E.~S. Toma, P.~Breger, G.~Mullot, F.~Aug\'{e}, Ph. Balcou, H.~G.
  Muller, and P.~Agostini.
\newblock Observation of a train of attosecond pulses from high harmonic
  generation.
\newblock {\em Science}, 292:1689, 2001.

\bibitem{Mairesse03}
Y.~Mairesse, A.~de~Bohan, L.~J. Frasinski, H.~Merdji, L.~C. Dinu,
  P.~Monchicourt, P.~Breger, M.~Kova\v{c}ev, R.~Ta\"{i}eb, B.~Carr\'{e}, H.~G.
  Muller, P.~Agostini, and P.~Sali\`{e}res.
\newblock Attosecond synchronization of high-harmonic soft x-rays.
\newblock {\em Science}, 302:1540, 2003.

\bibitem{Tzallas03}
P.~Tzallas, D.~Charalambidis, N.~A. Papadogiannis, K.~Witte, and G.~D.
  Tsakiris.
\newblock Direct observation of attosecond light bunching.
\newblock {\em Nature}, 426:267, 2003.

\bibitem{Hentschel01}
M.~Hentschel, R.~Kienberger, Ch. Spielmann, G.~A. Reider, N.~Milosevic,
  T.~Brabec, P.~B. Corkum, U.~Heinzmann, M.~Drescher, and F.~Krausz.
\newblock Attosecond metrology.
\newblock {\em Nature}, 414:509, 2001.

\bibitem{Drescher01}
Markus Drescher, Michael Hentschel, Reinhard Kienberger, Gabriel Tempea,
  Christian Spielmann, G.~A. Reider, P.~B. Corkum, and Ferenc Krausz.
\newblock X-ray pulses approaching the attosecond frontier.
\newblock {\em Science}, 291:1923, 2001.

\bibitem{Lein02}
M.~Lein, N.~Hay, R.~Velotta, J.~P. Marangos, and P.~L. Knight.
\newblock Interference effects in high-order harmonic generation with
  molecules.
\newblock {\em Phys. Rev. A}, 66:023805, 2002.

\bibitem{Kanai05}
T.~Kanai, S.~Minemoto, and H.~Sakai.
\newblock Quantum interference during high-order harmonic generation from
  aligned molecules.
\newblock {\em Nature}, 435:470, 2005.

\bibitem{Baker06}
S.~Baker, J.~S. Robinson, C.~A. Haworth, H.~Teng, R.~A. Smith, C.~C.
  Chiril\u{a}, M.~Lein, J.~W.~G. Tisch, and J.~P. Marangos.
\newblock Probing proton dynamics in molecules on an attosecond time scale.
\newblock {\em Science}, 312:424, 2006.

\bibitem{Lein07}
M.~Lein.
\newblock Molecular imaging using recolliding electrons.
\newblock {\em J. Phys. B}, 40:R135, 2007.

\bibitem{Itatani04}
J.~Itatani, J.~Levesque, D.~Zeidler, H.~Niikura, H.~P\'{e}pin, J.~C. Kieffer,
  P.~B. Corkum, and D.~M. Villeneuve.
\newblock Tomographic imaging of molecular orbitals.
\newblock {\em Nature}, 432:867, 2004.

\bibitem{vanderZwan07}
E.~V. van~der Zwan and M.~Lein.
\newblock Tomographic imaging of molecular orbitals in length and velocity
  form.
\newblock {\em AIP Conf. Proc.}, 963:570, 2007.

\bibitem{vanderZwan07-2}
E.~V. van~der Zwan, C.~C. Chiril\u{a}, and M.~Lein.
\newblock {\em to be submitted}.

\bibitem{Lewenstein94}
M.~Lewenstein, Ph. Balcou, M.~Yu. Ivanov, A.~L'Huillier, and P.~B. Corkum.
\newblock Theory of high-harmonic generation by low-frequency laser fields.
\newblock {\em Phys. Rev. A}, 49(3):2117, 1994.

\bibitem{Cao07}
Wei Cao, Peixiang Lu, Pengfei Lan, Weiyi Hong, and Xinlin Wang.
\newblock Control of quantum paths in high-order harmonic generation via a
  $\omega+3\omega$ bichromatic laser field.
\newblock {\em J. Phys. B}, 40:869, 2007.

\bibitem{Milosevic02}
D.~B. Milo\ifmmode \check{s}\else \v{s}\fi{}evi\ifmmode~\acute{c}\else
  \'{c}\fi{} and W.~Becker.
\newblock Role of long quantum orbits in high-order harmonic generation.
\newblock {\em Phys. Rev. A}, 66(6):063417, 2002.

\end{thebibliography}

\end{document}